\documentclass[letterpaper, twocolumn, superscriptaddress, showpacs, preprintnumbers, amsmath, amssymb,longbibliography, nofootinbib]{revtex4-1}

\usepackage{graphicx}
\usepackage{dcolumn}
\usepackage{bm}
\usepackage{color}
\usepackage{hyperref}
\usepackage{caption}
\usepackage{adjustbox}
\usepackage[lofdepth,lotdepth]{subfig}
\usepackage{caption}

\def\be{\begin{equation}}
\def\ee{\end{equation}}
\def\bea{\begin{eqnarray}}
\def\eea{\end{eqnarray}}

\def\bfr{\mathbf{r}}

\def\bfk{\mathbf{k}}

\newcommand{\ex}[1]{\left \langle #1 \right \rangle}
\newcommand{\sex}[1]{\langle #1 \rangle}

\begin{document}

\title{On the Hohenberg-Mermin-Wagner theorem and its limitations }  

\author {Bertrand I. Halperin}

\affiliation {Department of Physics, Harvard University, Cambridge MA 02138,~USA.}  

\begin{abstract}
Just over fifty years ago, Pierre Hohenberg developed a rigorous proof of the non-existence of long-range order in a two-dimensional superfluid or superconductor at finite temperatures. The proof was immediately extended by N. D.  Mermin and H. Wagner to the Heisenberg ferromagnet and antiferromagnet,  and shortly thereafter, by Mermin to prove the absence of translational long-range order in a two-dimensional crystal, whether in quantum or classical mechanics.   In this paper, we present an extension of the Hohenberg-Mermin-Wagner theorem to give a rigorous proof of the impossibility of long-range ferromagnetic order in an itinerant electron system without spin-orbit coupling or magnetic dipole interactions. We also comment on some situations where there are compelling arguments that long-range order is impossible but no rigorous proof has been given,  as well as situations, such as a magnet with long range interactions, or orientational order in a two-dimensional crystal, where long-range order can occur that breaks a continuous symmetry.  

\end{abstract}
\maketitle


\section{Introduction}
\label{sec:intro}

 In the fall of  1966, Pierre Hohenberg submitted an article to Physical Review in which he demonstrated a rigorous proof of the impossibliity of long-range order in a superfluid or superconductor at non-zero temperature, in either one or two dimensions. The paper was received by the journal on  October 24, but did not appear in print until June 1967.\cite{Hoh67} 
 
 A week before Pierre's submission, David Mermin and Herbert Wagner submitted a manuscript to Physical Review Letters, which contained a
 similar proof of the absence of long-range order in one and two dimensions for a  Heisenberg ferromagnet or antiferromagnet, with SU(2) symmetry, or a magnetic system with U(1) symmetry and an order parameter perpendicular to the symmetry axis.\cite{MW} 
The principal difference between the two papers is that Mermin and Wagner discussed spins on a lattice, whereas Hohenberg was concerned with bosons or fermions in the continuum.

Mermin and Wagner clearly state in their paper that they were aware of Pierre's earlier work, and that their own work was, in fact,  inspired by discussions with Pierre.  However, due to the rapid processing by  Physical Review Letters, the Mermin-Wagner  paper appeared in print in November, 1966, some six months before Pierre's. As a result, Pierre's precedence has been frequently ignored.  The two closely related proofs,
 which should be properly referred to collectively as the ``Hohenberg-Mermin-Wagner theorem",  have, unfortunately,  been very often cited in the literature simply as the ``Mermin-Wagner theorem".  
One can only hope that authors will be more careful in the future, and that Pierre's role will no longer be neglected.
 
 In the present article, written some fifty years later, I  make a few remarks about what  
 Hohenberg-Mermin-Wagner theorem (HMW) does and does not imply.  I shall also present a proof of an extension of the HMW theorem to rule out rigorously the possibility of  ferromagnetic order in a two-dimensional system of itinerant electrons, in the absence of terms in the Hamiltonian that explicitly break the SU(2) spin symmetry.

\section{Limitations of the rigorous theorem}

\subsection*{Phase transitions and quasi-long-range order}

Hohenberg, in his 1967 paper, was careful to emphasize that the absence of true long-range order did not rule out the possibility of superfluid transport at low temperatures, or of a sharp phase transition  as the temperature is raised. Furthermore, he was careful to note that there were already numerous non-rigorous arguments  against  the existence of long-range order in one-  and two-dimensional systems where such order would break a continuous symmetry,  dating as far back as the 1930s.\cite{Bloch,Peierls,Landau,Rice}  

At the time of Hohenberg's writing, it 
was already understood that if one assumes that the only important degree of freedom for a superfluid at low temperatures is the phase of the superfluid order parameter, one can conclude that the  pair correlation function 
$\sex{\psi^\dagger (\bf{r}) \psi (0)}$ should fall off as a power law, $ \propto r^{- \alpha}$ where $\alpha \to 0$ as $T \to 0$.\cite{Rice}
The key  assumption, here, is that the amplitude of the Bose condensate can be written as 
\be
a( \bfr ) = a_0 e^{i \theta ( \bfr )} ,
\ee
where  $a_0$ is treated as a constant while the phase $\theta$ is allowed to vary in space. Symmetry requires that  the energy is unchanged if the phase is changed by a constant independent of space, so the free energy cost of a spatial  variation in the phase should have the form
\be
\delta F =  \frac {\rho_s}{2} \int d^2 r | \nabla \theta |^2 ,
\ee
where $\rho_s $ is a finite stiffness constant. 
This leads to the result that at temperature $T$, phase fluctuations at wave vector $k$ should have a variance proportional to $T/(\rho_s k^2)$, which implies that the correlation function in real space,  $\sex { [\theta (\bf r )- \theta (0) ]^2 } $,  should diverge as  $(T/ \pi \rho_s) \log r$, for large $r$.   Because the phase fluctuations have a Gaussian distribution, this implies in turn that
\be
\ex { a( r ) a^* (0) } = e ^ {-  \sex { [\theta ( r )- \theta (0) ]^2 }/2 } \sim r^ { - T / (2 \pi \rho_s)}  .
\ee

It is now understood, following the work of Berezinskii,\cite{Berezinskii}  of Kosterlitz and Thouless,\cite{KosterlitzThouless,Kosterlitz} and of Nelson and Kosterlitz,\cite{NelsonKosterlitz}   that there will be a sharp transition temperature $T_{KT}$, such that for $T>T_{KT}$,  the correlation function falls off exponentially with distance, while for $T<T_{KT}$ the correlation function will have power law fall off, commonly called {\em quasi-long-range order}, with an exponent $\alpha \geq 3$.   By contrast, for the Heisenberg model, we now understand that the order-parameter correlation function  will decay exponentially with distance at any non-zero temperature, and there will generally be no  phase transition at finite temperatures.  The HMW theorem sheds no light on the question of whether quasi-long-range order can exist in any particular system.

\subsection*{Long-range interactions}

Hohenberg noted, in his original paper, that his theorem on the absence of long-range order for a two-dimensional superfluid or superconductor would be unaffected if the interaction between the particles included a long-range portion.  By contrast, the argument for the absence of long-range order in the Heisenberg ferromagnet depends crucially on the range of the interaction. In particular, let $J ({\bf{r}} )$ be the coupling constant between spin pairs separated by a distance $\bf{r}$,
and define second moments 
\be
K_{\alpha \beta} \equiv \sum_{\bfr}  r_\alpha r_\beta J ({\bfr}),
\ee
where $\alpha$ and $\beta$ denote the two spatial directions.  
The argument for the non-existence of long-range order,  as given in the Mermin-Wagner  paper requires that these second moments be finite.  

In the case of a magnetic model which has only  symmetry for spin rotations about one axis, say the z-axis, one must distinguish between coupling constants $J^{\perp}({\bf{r}})$ and $J^{\parallel}({\bf{r}})$ for spin components parallel or perpendicular to the symmetry axis, respectively.  Then, the analysis of Mermin and Wagner rules out  long-range order for spin components perpendicular to the symmetry axis,  provided that the second moments for $J^{\perp}$ are finite. It is not necessary to impose such a condition on $J^{\parallel}$.  

In fact, if the interactions are ferromagnetic and  fall off with distance slower than $1/r^4$, so that $K_{\alpha \alpha}$ is infinite, we expect that long-range ferromagnetic order should indeed be possible in two dimensions, at sufficiently small, but non-zero, temperatures.

\subsection*{Bond orientation in a two-dimensional crystal}

It has sometimes been suggested that the HMW theorem and its generalizations rule out the possibility of long-range order at $T \neq 0$ in any two-dimensional system where the order parameter would break a continuous symmetry.  However, this is not completely correct. We have already noted that long range order can occur in a Heisenberg magnet with suitably long-ranged interactions.  In a very different counter-example, one can have  long-range orientational order in a two-dimensional crystal, even if there are only short range interactions among the constituent particles.\cite{Mermin68}  

The possibility of long-range orientational order should be contrasted  with  translational order parameters, for which only quasi-long-range order is possible in two dimensions.  This distinction can be understood in a non-rigorous way by considering the effect of  thermally excited phonons, which will be present even in an ideal harmonic crystal, with no defects of any kind.  We may define a set of translational order parameters by
\be
\Psi_{\bf{G}}({\bf{r}}) = e^{i  {\bf{G}} \cdot {\bf{r}}   } [ \rho(\bfr) - \bar{\rho}] ,
\ee
where $\mathbf{G}$ is one of the fundamental reciprocal lattice vectors of the crystal,  $\rho(\bfr)$ is the particle density at point $\bfr$, and $\bar{\rho}$ is the average density.  
Then,  the effect of thermal phonons fluctuations in the long wavelength vibrational modes is to cause the order parameter correlations  to fall off at large distances as  
\be
\ex{ \Psi_{\bf{G}}({\bf{r}})  \Psi_{-\bf{G}}({\bf{0) }} }   \sim  r^{- T G^2 / \kappa },
\ee
where $\kappa$ depends on the elastic constants of the crystal.  
   
   If the crystal symmetry is such that each atom has $m$ nearest neighbors in equilibrium,  we may define an orientational order parameter as
\be
\Phi( \bfr ) =  \sex {e^ {i m \theta ( \bfr ) } }  _{ \rm {av} } ,
\ee
where $\theta$ is the orientation angle of a bond relative to the x-axis, and $ \sex {}_{ \rm {av}} $  denotes an average over the $m$  nearest neighbors of the atom at point $ \bfr $.  (For a simple triangular lattice, we have $m=6$.) 
Because long-wavelength phonons couple much more weakly to orientational order than to translational order,  phonons  by themselves do not prevent the existence of a non-zero expectation value  of $\sex {\Phi}$ independent of  $\bfr $.\cite{Mermin68}  If we denote the displacement field produced by a long-wavelength lattice distortion by 
$\mathbf{u}(\bfr)$, then the associated rotation will be given by $\delta \theta = \nabla \times \mathbf{u}$.  Consequently, the thermal fluctuations in $\theta$ will be smaller than the fluctuations in $\mathbf{u}$ by a factor of $k$  at long wavelengths. 

For a real crystal, one should consider the role of anharmonic terms and the possibility  of defects such as dislocations.  While the primary effect of anharmonic terms is only to renormalize the values of the elastic constants controlling the phonons at long wavelengths, the presence of un-paired dislocations  would  have a much larger effect than phonons on both orientational  and translational order.  However,   isolated dislocations have a logarithmically diverging energy in two dimensions, so dislocations should  only exist as bound pairs at low temperatures, which will only lead to a  small additional  renormalization of the elastic constants at low temperatures. 

It as been proposed that there can be a  temperature range where unbound dislocations exist, but free disclinations are not yet possible.\cite{NelsonHalperin} In this temperature range, the system will behave like a liquid crystal: correlations of the translational order parameters will fall off exponentially with distance, but there will be power law fall-off for correlations of the orientational order parameter.  This behavior of the orientational correlation function  may be understood because, in the  liquid-crystal phase, fluctuations in the orientational order have a free-energy cost proportional to the square of the gradient of $\Phi$, in contrast with the solid phase, where  a periodic fluctuation in $\Phi$ has a free-energy cost independent of the wave vector.  At the top of this temperature range, there would be a second transition, to an isotropic liquid phase where unpaired disclinations occur and there is exponential decay of correlations for both orientational and translational order.  Evidence for the existence of the liquid-crystal {\em hexatic} phase, for suitable forms of the inter-particle interaction, has been found in computer simulations and in experiments on colloidal particles.\cite{KapferKrauth,Gasser}

From a more fundamental point of view, the reason the HMW theorem cannot be used to rule out long range orientational order at low temperatures in this system 
 is that the rotation symmetry  broken by the long-range bond orientation involves a rotation in space of the positions of all atoms in the system. In such a rotation, the displacement of an atom will grow linearly with the distance from the origin. Consequently, the generator of these rotations cannot be written as an unweighted sum of local operators, as was the case for the symmetry generators in  the magnetic and superfluid systems. 
 
 \subsection*{Tethered Membranes}
 An interesting result for  orientational correlations has been found in a model  first proposed by Nelson and Pelliti in 1987, 
which  is a model for a suspended elastic membrane under zero tension, \cite{NelsonPeliti,NelsonBook} 
Consider the effects of thermal fluctuations about a ground state in which the membrane lies in the x-y plane. 
At least for small fluctuations, we may define the state of the system by specifying  a set of three functions,
$u_x(x,y), \, u_y(x,y), \, f(x,y)$, which describe, respectively, the in-plane and out-of-plane  displacements of the point originally at $(x,y,0)$.  For small displacements, the associated energy cost will be have the form
\be
\delta E \approx \frac{1}{2}  \int dx dy \left[  \kappa (\nabla^2 f)^2 + 2 \mu u_{ij} u_{ij} + \lambda u^2_{ii}      \right],
 \ee
 where $\kappa$ is a bending modulus,   $\mu$  and $\lambda$ are the in-plane elastic constants, and $u_{ij}$ is the strain tensor, given by
 \be
 \label{uij}
 u_{ij} \approx  \frac{1}{2} \left[ \partial_i u_j + \partial_j u_i + (\partial_i f) (\partial_j f) \right].
 \ee
 
 We shall be interested in fluctuations in the orientation of the membrane, defined by the local normal to the surface, $\hat{n}(x,y)$,   which may be written to lowest order in the displacements as
 \be
 \hat{n} = \hat{z} -\nabla f ,
 \ee
 where $\hat{z}$ is a unit vector in the z-direction.   If one were to neglect the terms involving the gradients of $f$ in Eq.~(\ref{uij}), there would be no coupling between the in-plane and out of plane displacements,  and fluctuations in $f$ would behave as $\sex { |f_\bfk|^2} \propto T / \kappa k^4$, for    $k \to 0$.  This would imply $\sex { |\delta \hat{n}|^2}  \propto T/\kappa k^2$ so that the mean square fluctuations in the orientation at a given point would diverge, for any $T>0$.
 However, the situation is changed when we take into account the non-linear terms in the energy that  result from coupling between $f$ and $u_i$.  These terms lead to a renormalization of the bending constant $\kappa$, which diverges for $k \to 0$,   and which removes the divergence of fluctuations in $\hat {n}$.  As a result, one  predicts that long-range orientational order will be preserved at low non-zero temperatures.\cite{NelsonBook}

\subsection*{Question of order at $T=0$}
The HMW theorem has nothing to say about the existence or non-existence of long-range order in a quantum mechanical system at $T=0.$ Yet, non-rigorous arguments similar to those invoked at finite temperatures, suggest that long-range order is  in fact impossible at T=0 in many circumstances. For example, it is strongly believed that  the superfluid order parameter of a one-dimensional collection of bosons  can have at most quasi-long-range order at $T=0$.\cite{Giamarchi}  If one makes the assumption that the spectrum of fluctuations at long wavelengths is dominated by a single phonon mode,  then one can easily make the argument, 
based on the quantum mechanical zero-point motion of the long-wave length phonons, that the phase fluctuations will diverge even  at T=0, ruling out the possibility of  long-range order.
As far as I am aware, however, there is no  generalization of the HMW theorem giving a {\em rigorous} proof of the absence of long range order in this case. 

A related situation is that of a two-dimensional fluid of bosons with a long-range  repulsive interaction.  Suppose that the repulsive force between two atoms  falls off at long distances proportional to $1/r^y$, with $y < 3$ .   (One needs to include a neutralizing uniform background in this case, in order to keep the energy density  finite and the particle density uniform.) The long-range repulsion leads to a phonon spectrum with  $\omega_k \propto k^{(y-1)/2} $, for $k \to 0$, and zero-point phase fluctuations  will have the form 
\be
\sex { | \theta_k | ^2} \propto   \,  \omega_k  k^{-2} \propto  \,   k^{(y-5)/2}  . 
\ee
If $y \leq 1$, the integral $\int d^2 k \sex { | \theta_k | ^2} $ will diverge, which implies that  long-range superfluid order is impossible, even at $T=0$.  Although this argument seems compelling, I am not aware of a rigorous proof that fluctuations must be dominated by a single phonon mode at long wavelengths, and I am not aware of a rigorous proof of the impossibility of long-range order in this case.

We remark that the borderline case $y=1$ is the situation  of a two-dimensional Coulomb interaction, were the interaction potential behaves as a logarithm of the distance.  In this case, the phonon (or plasmon)  frequency is non-zero and finite for $k \to 0$. For $y<1$, the frequency diverges for $k \to 0$.

\section{Extension of HMW to Itinerant ferromagnets}

The Mermin-Wagner paper discussed only the case of spins on a lattice.  However,  a generalization of the HMW argument  can  be used to rule out rigorously the possibility of ferromagnetism at $T \neq 0$ in any two-dimensional electron model without spin-orbit coupling or  magnetic dipole interactions, which would  destroy $SU(2)$ symmetry in  the microscopic Hamiltonian.  In particular, the model could employ any reasonable spin-independent two-body interaction and  it could include an 
an arbitrary periodic one-body potential.

The proof goes as follows.   As in the HMW papers, we make use of the inequality
\be
\label{FlucDis}
\ex {   \{ A , A^\dagger \} }  \geq   T \chi_{A, A} ,
\ee
where $ \chi_{A, A} $ is the linear response coefficient describing the expectation value of the operator $A$ produced by a static perturbation of the form $\lambda A +$ h.c.   The inequality follows directly from the quantum version of the fluctuation dissipation theorem.

Here, we consider a system of electrons with an unperturbed Hamiltonian of the form 
$H_0 = K + V - h S_x$, where the potential  $V$ is a spin-independent function of the positions of the particles, $K$ is the kinetic energy, $S_x$ is the x-component of the total spin,  and $h$ is a weak magnetic field that is allowed to vanish as the system size is taken to infinity.  We assume that the system is in a ferromagnetic state with a finite magnetization $s_x^0$, aligned in the  in the x-direction by the infinitesimal magnetic field $h$. We choose $A$ to be  $s_y (\bfk)$, the Fourier amplitude of the y-component of the spin density 
at wave vector $\bfk$.  

We now apply a weak perturbation of the form $( -\lambda A +$ h.c.$)$,  with $\lambda$ real and positive,  and investigate the energy change when the entropy is held fixed.   To lowest order in $\lambda$,  this will be given by 
\be
\label{Echi}
\delta E = - |\lambda|^2 \chi_{A,A}
\ee
We can establish a variational upper bound to  $\delta E$ by using a trial state where every eigenfunction of $H_0$ is multiplied by the unitary operator
\be
U = e^{- i  \eta Q} , 
\ee
where $\eta$ is a  real variational parameter, and 
\be
Q = 2 \int d^2 r \,s_z(\bfr)  \cos (\bfk \cdot \bfr)  =   [ s_z(\bfk) + s_z(-\bfk) ] .
\ee
 In the long wavelength limit, $U$ will produce a spatially-varying rotation of the state about the z-axis  by an angle 
 \be
 \theta(\bfr) = 2 \eta \cos (\bfk \cdot \bfr).
 \ee
   This will clearly lead to a non-zero value of $\sex{s_y}$, given by 
\be
\sex{s_y} = 2  s_x^0 \eta \cos (\bfk \cdot) \bfr .
\ee
The interaction of thiis magnetization with the applied perturbation will then give a contribution to the energy 
given by
\be
\delta E_1 =  - 2 \eta \lambda s_x^0   \Omega,
\ee
where $\Omega$ is the area of the system. 

The energy {\em cost} of the spatially-varying imposed rotation  will be given by 
\be
\delta E_2 = \ex  {U^{-1}  [H_0 , U ] } ,
\ee
where the expectation value is taken in the thermodynamic ground state of $H_0$. Since a uniform rotation costs no energy in the limit where the uniform field $h$ is taken to zero, we would expect that $\delta E_2$ will depend only on gradients of the rotation, Taking into account symmetries in the sign of $\eta$ and in the direction of $\bfk$, we would expect that the energy cost should be proportional to $\eta^2 k^2$.  Indeed we can calculate this energy cost precisely and verify this result.

Expanding  $U$ in powers of $\eta$, we find that to lowest non-vanishing order
\be
\delta E_2 =  \frac {\eta^2}{2} \ex  {Q  [H_0 , Q] } .
\ee
The operator $s_z (\bfr)$ is one-half the difference in densities of spin up and spin down particles at point $\bfr$,  so it clearly  commutes with the potential energy, which is a function of the density operators at various points. Consequently, we need only consider the commutator of $U$ with the kinetic energy $K$.  Following a similar procedure as is used in the standard derivation of the $f$-sum rule, one finds that  
\be
\delta E_2 =  \frac {\eta^2 k^2 n \Omega} { m }, .
\ee
where $n$ is the electron density and $m$ the electron mass.  By the variational principal, we know that $\delta E \leq \delta E_1 + \delta E_2$, so we get the best upper bound to $\delta E$ by choosing $\eta$ to minimize the right-hand side of this expression.  This gives 
\be
\delta E \leq   -  \frac { \lambda ^2 m \Omega}{n k^2}  (s_x^0)^2 ,
\ee
and, using (\ref{FlucDis}) and (\ref{Echi}),
\be
\label{sk2}
\ex { | s_y (\bfk) | ^2 } \geq \frac { T \Omega} {n k^2} (s_x^0)^2   .
\ee

We next define a locally averaged spin variable
\be
\bar{s} = \int d^2 r f(\bfr) s_y (\bfr )
 = \Omega^{-1} \sum_{\bfk} \tilde{f} (\bfk)  s_z(\bfk),
\ee
where $f$ is a gaussian with a finite  spatial width $a$,   and $\tilde{f}$ is the  Fourier transform of $f$.  In the limit of an infinite system, we have
\be 
\ex{ \bar{s}^2} = \frac {1} {(2 \pi)^2 \Omega } \int d^2k |\tilde{f}(\bfk)|^2   \ex { | s_y (\bfk) | ^2 }  ,
\ee
which, in view of Eq.~(\ref{sk2}), will be infinite if      $s_x^0 \neq 0$.
   But this is physically impossible.  A large value of $|\bar {s}|$ requires that the number of electrons $N_a$ in the region of size $a$ must be at least as large as $|\bar {s}|a^2 $.  But  for large densities, the kinetic energy per electron in the region will be at least of order $N_a / (ma^2)$. Therefore if $\sex{ \bar{s}^2}$ is infinite, the kinetic energy must also be infinite.  The Coulomb repulsion between electrons at short distances will only make things worse. 

It follows that the magnetization $s_x^0$ must be zero.

\section*{Note Added}

 Mohit Randeria has called my attention to a 1991 paper by Pitaevskii and Stringari \cite{PitaevskiiStringari}, which, in fact, gives a rigorous proof of the impossibility of superfluid long-range-order at zero temperature for a one-dimensional Bose system, subject to the assumption that the compressibility is finite at long wavelengths.  This paper also shows the impossibility of long-range crystalline order in a zero-temperature one-dimensional quantum system, and the impossibility of antiferromagnetic order in a one-dimensional quantum Heisenberg model, subject to the assumption of a finite value for the uniform magnetic susceptibility.

\section*{Acknowledgments}

My understanding of this subject has benefited greatly, over the years, from discussions with with many colleagues, but especially Pierre Hohenberg, Paul Martin, and David Nelson. I am grateful to  David Nelson also for helpful comments on the manuscript. 


\bibliography{Hohenberg-MW-Bib}

\end{document}